\documentclass[12pt]{article}
\usepackage{amsmath}
\usepackage{amscd}
\usepackage{amssymb}
\usepackage{array}

\begin{document}
\rightline{CERN-TH/2004-090}

\rightline{IFIC/04-20}

\rightline{FTUV-04/0518}

\newcommand{\R}{\mathbb{R}}
\newcommand{\C}{\mathbb{C}}
\newcommand{\Q}{\mathbb{Q}}
\newcommand{\Z}{\mathbb{Z}}
\newcommand{\Hb}{\mathbb{H}}
\newcommand{\Ss}{Scherk--Schwarz }
\newcommand{\KK}{ Kaluza--Klein }

\newcommand{\rE}{\mathrm{E}}
\newcommand{\ii}{\mathrm{i}}
\newcommand{\rSp}{\mathrm{Sp}}
\newcommand{\rSO}{\mathrm{SO}}
\newcommand{\rSL}{\mathrm{SL}}
\newcommand{\rSU}{\mathrm{SU}}
\newcommand{\rUSp}{\mathrm{USp}}
\newcommand{\rU}{\mathrm{U}}
\newcommand{\rF}{\mathrm{F}}
\newcommand{\rGL}{\mathrm{GL}}
\newcommand{\rG}{\mathrm{G}}
\newcommand{\rK}{\mathrm{K}}

\newcommand{\fgl}{\mathfrak{gl}}
\newcommand{\fu}{\mathfrak{u}}
\newcommand{\fsl}{\mathfrak{sl}}
\newcommand{\fsp}{\mathfrak{sp}}
\newcommand{\fusp}{\mathfrak{usp}}
\newcommand{\fsu}{\mathfrak{su}}
\newcommand{\fp}{\mathfrak{p}}
\newcommand{\fso}{\mathfrak{so}}
\newcommand{\fl}{\mathfrak{l}}
\newcommand{\fg}{\mathfrak{g}}
\newcommand{\fr}{\mathfrak{r}}
\newcommand{\fe}{\mathfrak{e}}
\newcommand{\ft}{\mathfrak{t}}
\newcommand{\id}{\relax{\rm 1\kern-.35em 1}}

\vskip 1cm

  \centerline{\LARGE \bf Scherk--Schwarz Reduction of $D=5$}

  \bigskip

   \centerline{\LARGE \bf Special and Quaternionic Geometry}

\vskip 1.5cm

\centerline{L. Andrianopoli$^{\flat}$,  S.
Ferrara$^{\flat,\sharp}$ and M. A. Lled\'o$^{\natural}$.}
 \vskip
1.5cm

\centerline{\it $^\flat$ CERN, Theory Division, CH 1211 Geneva 23,
Switzerland.} \centerline{{\footnotesize e-mail:
Laura.Andrianopoli@cern.ch, \; Sergio.Ferrara@cern.ch}}

\medskip

\centerline{\it \it $^\sharp$INFN, Laboratori Nazionali di
Frascati, Italy.}

\medskip
\centerline{\it $^\natural$
 Departamento de F\'{\i}sica Te\'orica,
Universidad de Valencia and IFIC}
 \centerline{\small\it C/Dr.
Moliner, 50, E-46100 Burjassot (Valencia), Spain.}
 \centerline{{\footnotesize e-mail: Maria.Lledo@ific.uv.es}}


\vskip 3cm

\begin{abstract}

We give the $N=2$ gauged supergravity interpretation of a generic $D=4$, $N=2$ theory as it comes from generalized Scherk--Schwarz reduction of 
$D=5$, $N=2$ (ungauged) supergravity.
We focus on the geometric aspects of the $D=4$ data such as the general form of the scalar potential and masses in terms of the gauging of a ``flat group''.
Higgs and super-Higgs mechanism are discussed in some detail.

\end{abstract}

 \vfill\eject


\section{Introduction}

New gaugings of extended supergravity have recently been considered, especially in connection to compactifications of M-theory and superstring theories 
involving generalized Scherk--Schwarz reduction \cite{ss} or flux vacua \cite{frey}.

In the present investigation we would like to give some general features of Scherk--Schwarz (SS) reduction of $D=5$, $N=2$ ungauged  supergravity \cite{gst}, in terms of what is generally
known for $D=4$, gauged supergravity \cite{dlv,n=2}.

This investigation can be regarded as the $N=2$ counterpart of a similar analysis \cite{adfl1} performed for the case of $N=8$ SS spontaneously
broken supergravity  \cite{css,svn}. $N=8$, $N=4$ and $N=2$ supergravity with SS reduction have been considered in the literature 
\cite{adfl2,dh,dst}, also
in connection with Randall--Sundrum brane scenarios \cite{rs,bhn,kl,dls,aq,gq,bfz,rss,pq,adpq}.

Following Scherk and Schwarz, we are going to consider in the following only {\em flat groups} obtained by SS reduction of a $D=5$ theory, thus getting, in $D=4$, 
a semi-positive definite scalar potential which generally has some flat directions, irrespectively if supersymmetry is broken or not.
Therefore the SS reduction always defines no-scale models \cite{ln}, which were first considered, in the context of $N=1$ supergravity, 
in the early 80's \cite{cfkn,eln}.
No-scale models also appear in warped superstring compactifications where ``internal'' fluxes are turned on \cite{tv,cklt,gkp,kst,fp}.

At the supergravity level the difference between \Ss and flux compactifications mainly resides in the fact that, when supersymmetry is broken,
massive multiplets are BPS saturated in the former case, while they are long in the latter.

The BPS $\rU(1)$ central charge is gauged by the graviphoton which, together with the other matter gauge fields, defines a {\em non-abelian} gauge
algebra corresponding to a {\em flat group}.

In $N=8$ supergravity the flat Lie algebra is a subalgebra of $\rE_{7(7)}$, while in $N=2$ it depends on the structure of the $D=5$ real-special geometry 
underlying the vector multiplet couplings \cite{adfl1}.

We find that the structure of the flat group gauged in $D=4$ is universal, in the sense that its structure is common to all $N=2$ models.
It is of the form $\rU(1)\circledS ~  \R^{n_v +1}$, where $n_v$ is the total number of $D=5$ vector multiplets, and in the most general case 
the $\rU(1)$ symmetry, which is gauged by the four-dimensional graviphoton $B_\mu$, 
\footnote{By graviphoton we call the four-dimensional vector coming from the five-dimensional metric.
This is not the same as the vector partner of the metric in the $N=2$, $D=4$ gravity supermultiplet.} acts on both the 
special manifold as well as the quaternionic manifold.

If the $\rU(1)$  has a component on the $\rSU(2)$ R-symmetry of the $D=5$ theory, then supersymmetry is broken.
When the quaternionic manifold is involved in the gauging, this is realized when the pull-back on space-time of the $\rSU(2)$ connection $\omega_A^{\ B}$ is non vanishing,
 with in particular
\footnote{Here and in the following we refer,
for the definitions of $D=4$, $N=2$ fields, to the ones  given  in \cite{n=2}.} 
$${\omega_5}_A^{\ B} \neq 0 \, ; \qquad (A,B=1,2) .$$
In absence of hypermultiplets, for supersymmetry to be broken the $\rU(1)$ must have a component on the global $\rSU(2)$ R-symmetry of the $D=5$ theory, and this 
originates a four-dimensional $N=2$ Fayet--Iliopoulos term. 

On the other hand,  if no SS phase is introduced in the real-special manifold, then the $\rU(1)$ has no component on the vector multiplet directions and
 the flat group is abelian, with all the vectors remaining massless.

 In the general case, the $\rU(1)$ charge has components on the isometries of  both  special and quaternionic manifolds.

We denote by $\Gamma^{ab}= -\Gamma^{ba}$ ($a,b =1,\cdots n_v$) and $\Delta^{\alpha\beta}=\Delta^{\beta\alpha} $  ($\alpha,\beta =1,\cdots 2 n_h$)
 respectively  the $\rSO(n_v)$ spin-connection one form on the real-special manifold 
and  the $\rUSp(2n_h)$ symplectic connection one-form  on the quaternionic manifold.
 For a Higgs mechanism to take place the following
conditions for the pull-back on 5D space-time of these connections must be met
$$ {\Gamma_5}^{a b}\neq 0 \,, \quad {\Delta_5}^{\alpha\beta}\neq 0\, . $$
 
Note that, if ${\omega_5}_A^{\ B}=0 $, then supersymmetry remains unbroken and a pure supersymmetric Higgs mechanism occurs.  
Then massive BPS multiplets are generated for both vector- and hypermultiplets in four dimensions.

In Section 2 we discuss the general form of the scalar potential and the fermionic bilinear of a generic compactification, as they come from $D=5$ SS reduction.

In Section 3 we give the basic gauge groups and discuss the SS reduction in terms of $D=4$, $N=2$ data based on the gauging of a ``flat group''.

In Section 4 we apply this general pattern to different models, covering the different cases with $n_v, n_h =0$; $n_v\neq 0, n_h =0$; 
$n_v=0, n_h \neq 0$; $n_v, n_h \neq 0$. We mostly confine our analysis to symmetric coset spaces.

We give in two appendices further relations between SS $D=5$ dimensionally reduced quantities and their $D=4$ gauged supergravity counterparts.

Further extensions of the present investigation, which will be considered elsewhere, include the presence of a more general lagrangian in $D=5$, as, for example, a
non-abelian gauge group already in $D=5$, opening the possibility of having a $D=4$ boundary, as in the Randall--Sundrum kind of scenario, or 
 the inclusion of tensor multiplets in $D=5$ \cite{gzag,cd,sorin}, through which one would retrieve in $D=4$ a more general $N=2$ matter coupling discussed in 
reference \cite{ddsv}.  


\section{On the \Ss reduction of the $D=5$, $N=2$ lagrangian}

The basic quantities of the $D=5$ lagrangian which become relevant in the discussion of the SS reduction are the kinetic terms for the scalars 
and for the spin $\frac 12 , \frac 32 $ fields, which are related to the scalar potential and the fermionic mass terms of the $D=4$ dimensionally reduced theory.

These kinetic terms contain, in the fermionic covariant derivatives, the real-special geometry spin-connection \cite{dvvp}, 
as well as the symplectic $\rUSp(2n_h)$ and $\rSU(2)$ connections of the quaternionic manifold \cite{bw,n=2}.

Upon SS reduction, the pull-back on space-time of such one-form connections contribute the terms
$$ {\Delta_5}_{\alpha\beta}\, ;  \quad {\omega_5}_{AB}\, ; \quad {\Gamma_5}^{ab} \qquad\qquad\qquad ({\omega_5}_{AB}={\omega_5}_{A}^{\ C} \epsilon_{BC}) $$ 
which will determine the quadratic mass terms of the fermions.

Similarly, the $D=4$ scalar potential is \cite{svn}
\begin{equation}
V(\sigma , \varphi , q)=  e^{-3\sigma}\left[\frac 12 P_5^a(\varphi) P_{5 a}(\varphi) + {\mathcal {U}}_5^{\alpha A}(q)\,{\mathcal {U}}_5^{\beta B}(q)\C_{\alpha\beta}
 \epsilon_{AB}\right]
\label{5dpot}
\end{equation}
where $\sqrt{g_{55}}=e^\sigma=\phi_5$, $P_5^a(\varphi) = P^a_{i} \partial_5 \varphi^{i}$ and ${\mathcal {U}}_5^{\alpha A}(q)= {\mathcal {U}}_u^{\alpha A} \partial_5 q^u$ denote the 5th components
 of the pull-back on space-time of the scalar vielbeins of the theory in $D=5$, with $P^a_{i}$ the vielbein of the real-special manifold in $D=5$ and 
${\mathcal {U}}_u^{\alpha A}$ the vielbein of the quaternionic manifold (with $i=a=1,\cdots n_v$; $A=1,2$; $\alpha = 1,\cdots 2n_h$ and $u=1,\cdots 4n_h$). 
$\phi_5=e^\sigma =\sqrt{ g_{55}}$ is the \KK mode from the metric.

It is obvious from (\ref{5dpot}) that the potential $V$ is positive definite and it has an extremum at the points for which
\begin{equation}
P_5^a(\varphi) ={\mathcal {U}}_5^{\alpha A}(q)=0 .
\end{equation}
These are the vacua of the theory.

As we already know from the $N=8$ example, these vacua may have a non trivial moduli space (other than the $\sigma$ direction).
Also, the scalar potential (\ref{5dpot}) can be recast in a $N=2$, $D=4$ gauged supergravity form, by giving 
the gauged isometries which correspond to the SS compactification.

It is important to observe that, in view of the real-special geometry relations discussed in Appendix A \cite{cfgk,ccdf}, 
the first term in the potential can also be written as 
\begin{equation}
 -\frac 32 e^{-3\sigma} \partial_5 t^I \partial_5 t^J t^K d_{IJK}
\label{veryspec}
\end{equation}
where $t^I(\varphi)$ ($I=1,\cdots n_v+1$) are  $D=5$ special coordinates, subject to the constraint 
\begin{equation}
t^I t^J t^K d_{IJK}=1 .
\label{constr}
\end{equation}
 They form a representation of the
full duality group of the $D=5$ theory.
 
Let us consider the term $\partial_5 t^I(\varphi) = t^I_{,i} \partial_5 \varphi^i$.
When applying the SS generalized dimensional reduction it gives, at $x^5=0$, 
\begin{equation}
\partial_5 t^I(\varphi)|_{x^5=0} = t^I_{,i} M^i_j \varphi^j ,
\label{ssphase}
\end{equation}
 where $M^i_j$ is a matrix parametrizing  the global symmetries of the $D=5$ action.
We restrict  our discussion to the case where $M^i_j$ only contains the compact symmetries.
In a more general sense, $\partial_5 \varphi^i=k^i_0$ is the Killing vector for an arbitrary $\rU(1)$ in the Cartan subalgebra of the global compact 
symmetry $H_\R$ of real-special geometry,
it is therefore a generator $\rU(1)$ depending on $n=$ rank $(H_\R)$ parameters. 

With the position (\ref{ssphase}) the vector multiplet sector of the $D=4$ scalar potential then becomes
\begin{equation}
\frac 12 e^{-3\sigma} g_{ij} M^i_k M^j_\ell \varphi^k \varphi^\ell
\label{5dvmpot}
\end{equation}
where (see Appendix B)
\begin{equation}
g_{ij} =-3 d_{IJK}t^K t^I_{,i} t^J_{,j}= G_{IJ}t^I_{,i}t^J_{,j}
\end{equation}
is the real-special manifold metric.

We are interested in comparing the above expression with the $D=4$ gauged supergravity result.


\section{Gauged $N=2$, $D=4$ interpretation}


\subsection{Vector multiplet sector}
In the $D=4$ framework, mass terms and scalar potential arise from the gauging procedure.
This was fully exploited for the $N=8$ case in \cite{adfl1}, and the same procedure was indicated for the $N=2$ case in Section 4 of the same reference.

The crucial ingredient to be used here is that the dimensional reduction of real-special geometry \cite{gst} gives a cubic holomorphic prepotential 
at $D=4$ of the form
\begin{equation}
F(z) = d_{IJK}z^Iz^Jz^K
\label{cubicprep}
\end{equation}
where $\Im z^I =v^I = t^I \phi_5 $, $ \Re z^I =- A^I_5$, with $A^I_5$ being the 5th components of the $D=5$ vectors. The special K\"ahler manifolds
with the  specific form (\ref{cubicprep}) for the prepotential are named {\em very-special } manifolds \cite{vanp} (or {\em d-manifolds}) \cite{dvvp}
\footnote{A d-manifold can be real (at $D=5$, with $n_\R =n_v$), K\"ahler (at $D=4$, with $n_\C =n_v +1$), or quaternionic (at $D=3$, with $n_h =n_v+2$),
 where $n_\R$, $n_\C$ and $n_h$ denote the real, complex and quaternionic dimension respectively. }.

Note that the $D=4$ moduli $v^I$ (corresponding to the imaginary part of the scalars in the vector multiplets) obey the relation
\begin{equation}
\mathcal{V} (v) \equiv d_{IJK}v^I v^J v^K = e^{3\sigma}
\label{radius}
\end{equation}
which immediately follows from equation (\ref{constr}) by using $v^I = t^I e^\sigma$.

Moreover, since the quantity $g_{ij}\varphi^k\varphi^\ell$ in (\ref{5dvmpot}) is scale invariant, we can read the factor $e^{-3\sigma}$ as 
\begin{equation}
e^{-3\sigma}=e^K
\label{kahlerphi}
\end{equation}
which defines the very-special K\"ahler potential $K= - \log \mathcal{V}= -3 \log \sigma$.

The $D=4$ vectors are $B_\mu$ (the \KK graviphoton) and $Z^I_\mu = A^I_\mu -A^I_5 B_\mu$.
This in particular shows that the five-dimensional graviphoton belongs, together with $\phi_5 =\sqrt{g_{55}}$, to an additional vector multiplet in four dimensions, 
while the $D=4$ graviphoton comes from the \KK vector, corresponding to the decomposition of the five-dimensional  space-time vielbein as
\begin{equation}
\left( \hat V^a_\mu = e^{-\frac\sigma 2} V^a_\mu ; \quad  \hat V^5_\mu = e^{ \sigma }  B_\mu  ; \quad  \hat V^5_5 = e^{ \sigma} \right)
.\end{equation}
This is merely due to the choice of $D=4$ special coordinates $X^\Lambda$ (see Appendix A) which set $X^0$ to correspond to $B_\mu$. This is not 
the same as the ``free'' supermultiplet assignment.

The very-special manifold $\mathcal{M}_{\C}$ in $D=4$ has $\R^{n_v+1}$ isometries (corresponding to the $A^I_5$ shift invariance of the lagrangian)
\begin{equation}
\delta A_5^I = r^I \, , \qquad I=1,\cdots n_v+1.
\end{equation}
They act on the $n_v+2$ vectors in $D=4$ as follows
\begin{equation}
\delta Z^I_\mu = - r^I B_\mu \, ; \quad \delta B_\mu =0 .
\end{equation}

Let us now consider $B_\mu $ to gauge a $\rU(1)$ group belonging  to the maximal compact subgroup of the isometry group of the very-special  manifold, 
and in particular let us take it
 in its Cartan subalgebra $\mathcal{H}_{\C}$.

If $t^I$ is a representation of $\mathcal{H}_{\C}$, then we may consider the following flat group
\begin{eqnarray}
&\left[ t^I, t^0 \right] & =
\  M^I_J t^J \nonumber\\
&\left[ t^I,t^J \right] & =
\ 0
\end{eqnarray}
that is, by setting $t^\Lambda =(t^0, t^I)$ ,
\begin{equation}
\left[ t^\Lambda, t^\Sigma \right]
= f^{\Lambda\Sigma}_{\ \  \Delta} t^\Delta\, , \qquad \Lambda =(0,I)
\end{equation}
with $f^{I0}_{\ \ J} = M^I_J$, the others vanishing.
 The  $\rU(1)$ isometry gauged by the $B_\mu$ gauge field may have components 
both on the very-special manifold and on the quaternionic manifold, that is \footnote{By $I^Q_{\rUSp(2n_h)}$ and  $I^Q_{\rSU(2)}$ we mean the SS phase with
flattened indices \cite{svn}.}
$$t^0=I^{SG} + I^Q_{\rUSp(2n_h)} + I^Q_{\rSU(2)}.$$ 
The charges corresponding to gauging some special geometry isometries $I^{SG}$  are given by the SS phase $M^I_J$,
while the ones corresponding to gauging quaternionic isometries $I^Q$ are labeled by a matrix $M^u_{\ v}$ which will be better specified in Section 3.3
(see in particular equation (\ref{massquat})).

The gauge transformation of the $z^I$ coordinates is holomorphic and has the form
\begin{equation}
\delta z^I = M^I_J \left(z^J \xi^0 - \xi^J\right)
\label{gaugetrscalars}
\end{equation} 
Of course, as the gauge parameters $\xi^\Lambda$ are real, the non-homogeneous part only affects $\Re z^I$.

From the structure constants of the non abelian gauge algebra we also get the gauge transformation of the vectors
\begin{eqnarray}
\delta Z^I_{\mu} &=& \partial_\mu \xi^I + M^I_J \left(\xi^0 Z^J_\mu - \xi^J B_\mu \right)\nonumber\\
\delta B_{\mu} &=& 
\partial_\mu \xi^0
\label{gaugetrvectors}
\end{eqnarray}
and the expression  for the vector field-strengths
\begin{eqnarray}
F^I_{\mu\nu} &=&  \partial_\mu Z^I_\nu -\partial_\nu Z^I_\mu + M^I_J \left(Z^J_\mu B_\nu - Z^J_\nu B_\mu \right)\nonumber\\
B_{\mu\nu} &=& 
\partial_\mu B_\nu -\partial_\nu B_\mu 
.\end{eqnarray}
Note that, because of the gauged translations, a Chern--Simons-like term is present in the $D=4$ lagrangian \cite{dlv}
\begin{equation}
\frac 13 \epsilon^{\mu\nu\rho\sigma} d_{IJK} M^K_L Z^I_\mu Z^L_\nu \partial_\rho Z^J_\sigma .
\end{equation}
It comes by dimensional reduction of the $D=5$ Chern--Simons term \cite{adfl1}
\begin{equation}
d_{IJK} A^I \wedge F^J \wedge F^K .
\end{equation}
Note that $d_{(IJK} M^K_{\ L)}=0$ and that the non-abelian contributions vanish identically \cite{adfl1}.

\bigskip

In $D=4$ very-special geometry, as it comes from dimensional reduction of $D=5$ real-special geometry, 
the choice of symplectic basis is dictated by the fact that the $D=5$ duality group does not mix electric with magnetic vector potentials, so that the flat group
is entirely electric. For infinitesimal transformations it has  then  a symplectic $\rSp(2n_v+4,\R)$ action \cite{gz} of the type
\begin{equation}
\begin{pmatrix}
a&0\cr c&-a^T\cr\end{pmatrix} \qquad (\mbox{with }c=c^T)
\label{symplaction}
\end{equation}
In (\ref{symplaction}), the submatrix $a$ corresponds to the adjoint action on the vectors, while $c$ is the matrix which transforms, non linearly,
the axion fields.
Setting all together, the entries of $a$, $c$ for the $(n_v+2)$-dimensional gauge algebra are
\begin{eqnarray}
a_{(n_v+2) \times (n_v+2)}&=&
\begin{pmatrix}
M^I_J \xi^0 & - M^I_J \xi^J \cr
0_{(n_v+1) \times (n_v+1)} & 0\cr
\end{pmatrix}
, \nonumber\\
c_{(n_v+2) \times (n_v+2)}&=&
\begin{pmatrix}
d_{IJK}M^K_L \xi^L & 0_{(n_v+1) \times 1} \cr
0_{(n_v+1) \times (n_v+1)} & 0\cr
\end{pmatrix}
\end{eqnarray}
In particular, $c$ determines the non-homogeneous shift for the vector kinetic matrix of very-special geometry (which can be computed with the
standard special-geometry formulae)
\footnote{Note that such possibility extends the results of \cite{n=2}, following \cite{dlv}.}
$$
\delta \Re \mathcal{N}_{IJ}=d_{IJK}M^K_L \xi^L
$$
which is related to the gauged axion symmetry
$$\delta z^I = - M^I_J \xi^J.$$


\subsection{Killing vectors and the scalar potential}

The $N=2$ scalar potential for a generic gauged supergravity is given by the following expression \cite{n=2}
\begin{equation}
V=\left(g_{I\bar J}k^I_\Lambda k^{\bar J}_\Sigma + 4 h_{uv} k^u_\Lambda k^v_\Sigma\right) \bar L^\Lambda L^\Sigma +
\left(U^{\Lambda\Sigma}-3 \bar L^\Lambda L^\Sigma\right)P^x_\Lambda P^x_\Sigma
\end{equation}
Here  $k^I_\Lambda$, $k^u_\Lambda$ are the Killing vectors of the special and quaternionic manifold respectively, $g_{I\bar J}$, $h_{uv}$ the 
corresponding metrics on the two manifolds,
 $L^\Lambda$ is the covariantly holomorphic section of special geometry, 
$P^x_\Lambda$ is the quaternionic Killing-prepotential, and $U^{\Lambda\Sigma}=\mathcal{D}_I L^\Lambda \mathcal{D}_{\bar J}\bar L^\Sigma g^{I\bar J}
= -\frac 12 (\Im \mathcal{N})^{-1 \Lambda\Sigma} -  \bar L^\Lambda L^\Sigma$.

We refer to \cite{n=2} for the complete definitions; however, 
note in particular the relations $g_{I\bar J} k^{\bar J}_\Sigma = \partial_I P_\Lambda$, $k^I_\Lambda L^\Lambda =P_\Lambda L^\Lambda =0$ 
($P_\Lambda$ is the special-K\"ahler Killing prepotential)
and $\Omega^x_{uv} k^v_\Lambda = \mathcal{D}_u P^x_\Lambda$, $  P^x_\Lambda = \frac 1{2n_h}\mathcal{D}_u  k^v_\Lambda (\Omega^x)^u_{\ v}$.

We note
that, for the kind of gauging under consideration, 
 the scalar potential, as expected by comparison with (\ref{5dpot}), is positive semidefinite.
This is a consequence of the fact that in these models only compact quaternionic isometries, corresponding to a SS phase, are  gauged, with the $\rU(1)$ graviphoton 
 as gauge vector, so that $P^x_0 \neq 0$, $P^x_I=0$.
Then, in the special coordinates, where
$$L^0 = e^{\frac K 2} ;  L^I =  e^{\frac K 2}z^I,$$
one can prove the basic relation \cite{ckpdfwg,gst2}
  $$U^{00} - 3  \bar L^0 L^0 \equiv 0,$$ so that the scalar potential reduces to
\begin{equation}
V=  e^K  \left(g_{I\bar J}k^I_\Lambda k^{\bar J}_\Sigma + 4 h_{uv} k^u_\Lambda k^v_\Sigma\right) \bar z^\Lambda z^\Sigma .
\label{positivepot}
\end{equation}

The two contributions in (\ref{positivepot}) have now to be compared with the two terms in (\ref{5dpot}).

\paragraph{No hypermultiplets\\}

Let us first consider the case $k^u_\Lambda =0$, where the graviphoton has no components on any $\rU(1)$ isometry of the quaternionic manifold.
First of all, from (\ref{gaugetrscalars}) we know the form of the very-special manifold Killing vectors:
\begin{equation}
k^I_J = -M^I_J \, ; \quad k^I_0 = M^I_J z^J 
\label{smkilling}
\end{equation}
which implies that the following properties of special geometry are true
\begin{eqnarray}
k^I_\Lambda L^\Lambda &=& e^{\frac K2}\left( k^I_J z^J + k^I_0\right) =0 \\
k^I_\Lambda \bar L^\Lambda &=& 2{\rm i}e^{\frac K2} M^I_J \Im z^J . \label{klb}
\end{eqnarray}
Equation (\ref{klb}) is the basic identity which allows to relate such an expression to $\partial_5 t^i$, as defined in (\ref{ssphase}).

Then the scalar potential is independent of $\Re z^I$
\begin{equation}
V(\sigma,\varphi)= e^K g_{I\bar J} k^I_\Lambda k^{\bar J}_\Sigma \bar z^\Lambda  z^\Sigma = 2 e^K G_{IJ}M^I_K M^J_L \Im z^K  \Im z^L
.\label{sgpot}
\end{equation}
Note that, when the set of indices $I,J...=1,\cdots n_v +1$ is restricted to the subset $i,j,...=1,\cdots n_v$, then (\ref{sgpot}) does coincide with (\ref{5dvmpot}).

The extrema are at 
$$ 
M^I_J \Im z^J =0, $$
that is they correspond to the vanishing eigenvalues of the Cartan matrix $M^I_J$.

Since the matrix $M^I_J$ is taken on a compact subgroup of the isometry group, it is anti-hermitean, with imaginary  eigenvalues $(\pm {\rm i} m_i,0)$.
The non-vanishing eigenvalues provide massive BPS $N=2$ vector multiplets, where $\Im z^I$ is the scalar partner of the massive vector $Z^I_\mu$ and 
$\Re z^I$ the corresponding would-be Goldstone boson.

The vanishing eigenvalues give the moduli space of the SS theory. If supersymmetry is unbroken ($ P^x_\Lambda =0$), they exactly correspond to
the number of residual massless vector multiplets. 

Let us consider now a similar situation, without hypermultiplets, but with a non-zero Fayet--Iliopoulos term $P^x_\Lambda = $ const.
As for an ordinary quaternionic prepotential, it must satisfy the constraint \cite{n=2}
$$-\frac 12 \epsilon^{xyz} P^y_I P^z_0 + \frac 12 f_{I0}^{\ \ J}P^x_J =0$$
which is solved for 
$P^x_I=0 $, $P^x_0$ arbitrary.
Let us then set $P^{x=3}_0 = m \neq 0$.
Supersymmetry is broken, but the vacuum is the same as before, since $m$ does not contribute to the scalar potential. However, it gives a mass term to the gravitinos
and to the gauginos.

This is a no-scale model with $N=2 \to N=0$ broken supersymmetry.
The two fermionic partners of the matter vector multiplet which was not present in $D=5$ ($\psi_{5 A}$ component of the gravitino) 
are the goldstinos which make the gravitinos massive.
In this case, the $\rU(1)$ gauged by the $B_\mu$ field is a combination of the special-manifold isometry and of the R-symmetry.

We note that if in this model we set to zero the matrix $M^I_J$, we retrieve the vanishing potential model discussed long-ago in \cite{ckpdfwg}.

\paragraph{Inclusion of hypermultiplets\\}
The most general case of SS gauging is obtained when the $B_\mu$ field has also a component on a $\rU(1)$ isometry of the quaternionic manifold.
In this case the SS phase takes values also on the isometries of the quaternionic geometry.

The scalar potential is given by (\ref{5dpot}) or, in the $D=4$ language, by (\ref{positivepot}).
Moreover, since on the quaternionic manifold $k^u_0 \neq 0$ is the only non-vanishing Killing vector, 
also the hypermultiplet contribution to the scalar potential is  independent from the axions in the vector multiplets (as expected), and is given by
\begin{equation}
4  e^K   h_{uv} k^u_0 k^v_0
\label{4dhmpot}
\end{equation}
where the factor $ e^K $ correctly reproduces the radial factor $e^{-3\sigma}$, as discussed in (\ref{radius}).

The flat vacuum corresponds to $k^u_0=0$.

From the hypermultiplet $\sigma$-model we have that the five-dimensional kinetic lagrangian which originates the scalar potential is nothing but
\begin{equation}
h_{uv} \partial_5 q^u \partial_5 q^v \label{kinpot}
\end{equation}
which, assuming a non-trivial SS phase in the quaternionic direction
$$\partial_5 q^u = M^u_{\ p} q^p\, ; \qquad (M^u_{\ p}=-M_p^{\ u})$$ 
becomes 
\begin{equation}
  h_{uv} M^u_{\ p} M^v_{\ r} q^pq^r 
.\label{5dhmpot}
\end{equation}
Comparing (\ref{4dhmpot}) with (\ref{5dhmpot}), and sending $M \to 2M$ in the SS phases, we have the identification
\begin{equation}
k^u_0 = M^u_{\ p} q^p \label{k0}
\end{equation}
with, in the general case where the SS phase takes value on all the $\rSU(2) \times \rUSp(2n_h)$ holonomy,
\begin{equation}M^u_{\ p}= \mathcal{U}_{\alpha A}^u \mathcal{U}_{p|\beta B} \left[ M_{AB} \C_{\alpha\beta} + 
M_{ \alpha\beta} \epsilon_{AB}
\right] 
\label{massquat}
\end{equation}
where $M_{AB} = M_{BA}$, $M_{\alpha\beta}=M_{\beta\alpha}$.

The flat directions of $k^u_0=0$ correspond to the vanishing eigenvalues of the matrix $M^u_{\ p}$.
The hypermultiplet scalars which become massive belong to massive BPS hypermultiplets
\footnote{Note that a massive hypermultiplet (unlike the vector multiplets) is necessarily BPS saturated.}.


\subsection{Mass terms and critical points of the scalar potential}

The strategy for finding mass terms in fermion bilinears, as they come from the five-dimensional theory compactified \`a la Scherk--Schwarz, 
 is to consider all terms in the lagrangian  which include either $P_5$ and $\omega_5$
(the 5th component on space-time respectively of the vielbein and of the spin-connection on the scalar manifolds).

Let us remind the reader that in $D=5$ the vector multiplet geometry has a Riemannian connection $\Gamma_i^{\ j}$ while the hypergeometry has 
$\rSU(2) \times \rUSp(2n_h)$ holonomy, with connections $\omega_A^{\ B}$ and $\Delta_\alpha^{\ \beta}$ respectively.
We denote the pull-back on space-time of the corresponding scalar vielbein as $P^a_{\hat \mu}=P^a_j \partial_{\hat \mu} \varphi^j $ 
(such that $P^a_j P_{a k}=g_{ik}$, with $\hat \mu =0,1,2,3,5=(\mu,5)$) and
$\mathcal{U}^{A\alpha}_{\hat \mu}= \mathcal{U}^{A\alpha}_u \partial_{\hat \mu}q^u$ ($\mathcal{U}^{A\alpha}_u \mathcal{U}^{B\beta}_v \epsilon_{AB} \C_{\alpha\beta} = h_{uv}$).
For what concerns the quaternionic vielbein $\mathcal{U}^{\alpha A}_{\hat \mu}$,
under SS dimensional reduction we have
$$\mathcal{U}^{\alpha A}_{5}=\mathcal{U}^{\alpha A}_{u}\partial_5 q^u = \mathcal{U}^{\alpha A}_{u} k^u_0$$
where $$k^u_0=\partial_5 q^u = m^u_{\ v} q^v$$ is the Killing vector corresponding to a $\rU(1)$ isometry with components on both symplectic and $\rSU(2)$ indices.

The contribution to the scalar potential from the quaternionic sector of the SS reduced lagrangian is
\begin{eqnarray*}
V_{Q}&=& e^{-3\sigma} \mathcal{U}^{\alpha A}_5 \mathcal{U}^{\beta B}_5 \C_{\alpha\beta} \epsilon_{AB}
\\
&=&  e^{-3\sigma} \mathcal{U}^{\alpha A}_u \mathcal{U}^{\beta B}_v k^u_0 k^v_0 \C_{\alpha\beta} \epsilon_{AB}
\\
&=&  e^{-3\sigma} h_{uv} k^u_0 k^v_0 
\end{eqnarray*}
which is indeed the $N=2$ expression for a $\rU(1)$ isometry on the quaternionic manifold gauged by the graviphoton.

By the general discussion on SS dimensional reduction we know that, if a SS phase in the R-symmetry direction is switched on (so that, in the $N=2$ cases considered here,
 ${\omega_5}_A^{\ B} \neq 0$) 
then supersymmetry is spontaneously broken.
Note that from the five-dimensional gravitino kinetic term we can read-off its four-dimensional bilinear
$$S^{AB} \bar\psi_{A\mu} \gamma^{\mu\nu} \psi_{B\nu}$$
if we interpret, in analogy to \cite{svn},
\begin{equation}
S_{AB} = \frac \ii 2 e^{-\frac 32 \sigma} (\omega_5)_{AB} .\label{gravmass5d}
\end{equation}
Comparing (\ref{gravmass5d}) with the general four-dimensional expression for $S_{AB}$, which in our case is
\begin{equation}
S_{AB} =\frac \ii 2 P_\Lambda^x (\sigma^x)_{AB} L^\Lambda =   P_0^{x} (\sigma^x)_{AB} e^{\frac K2}\label{gravmass4d}
\end{equation}
we get, recalling (\ref{kahlerphi}), 
\begin{equation}
  P_0^{x}=  \ii \omega_5^{x} . \label{gravmasscomb}
\end{equation}

Making use of the general formulae for quaternionic geometry, we can find an explicit expression for the Killing prepotential $P^{AB}_0$ 
which gives a mass term to the gravitinos.
Let us recall in particular  the relations:
\begin{eqnarray}
\Omega^x_{uv} &\equiv & {\rm i}  \mathcal{U}^{\alpha A}_u \mathcal{U}^{\beta B}_v \C_{\alpha\beta}(\sigma^x)_{AB}\, , \label{qg1}\\
\Omega^x_{uv}{\Omega^y}^v_{p}&=& - h_{up}\delta^{xy} - \epsilon^{xyz} \Omega^z_{up}\label{qg2}
\end{eqnarray}
and 
\begin{equation}
P^x_{\Lambda} = -\frac 1{2n_h}\nabla^u k^v_\Lambda \Omega^x_{uv}. \label{plam}
\end{equation}
From the analysis of the previous subsection, we know that a SS-induced quaternionic  Killing vector must be of the form (\ref{k0}).
In order for it to break supersymmetry, the mass matrix $M^u_{\ p}$ (\ref{massquat}) must have a non trivial phase on the $\rSU(2)$ isometry ($M_{AB} \neq 0$), that is
$$M^u_{\ p} = M^x(\sigma^x)^{AB}  \C^{\alpha\beta} \mathcal{U}_{\alpha A}^u \mathcal{U}_{\beta B |p}=- {\rm i} M^x{\Omega^x}^u_{\ p} $$
with $M^{AB} \equiv \frac 12 M^x(\sigma^x)^{AB}$, 
so that  (\ref{plam}) gives, for our case
\begin{equation}
P^x_0 \to -\frac 1{2n_h}\nabla^u k^v_0 \Omega^x_{uv}
= \frac {\rm i}{2n_h} M^y\Omega^x_{uv}{\Omega^y}^v_{\ p}\nabla^u q^p 
. \label{p0}
\end{equation}

Using now
$\nabla_u q^v = \delta_u^v + \mathcal{O}(q)$ and (\ref{qg2}) we have, for $q^u \to 0$,
\begin{equation}
P^x_0 =-2{\rm i} M^x \label{p0final}
\end{equation}
in accordance with the fact that, for $q^u \to 0$, $\omega_5^{AB} \to M^{AB}$.

We therefore see as expected that in absence of hypermultiplets the $M^{AB}$ parameter is the Fayet--Iliopoulos term,
while in presence of hypermultiplets it is related to the Killing vector
$$k^u_0 =- {\rm i} M^x(\Omega^x)^u_{\ v}q^v$$
corresponding to a $\rU(1) \subset \rSU(2)$ quaternionic isometry gauged by the graviphoton.

From equations (\ref{kinpot}) and (\ref{5dhmpot})
we see that in this case a mass term for the hypermultiplet scalars appears which stabilizes all of them, since, from (\ref{k0}),
 the extrema of the potential are found  here for $q^u=0$. 

The presence of a Fayet-Iliopoulos term is understood, from the SS point of view, from the fact that (see (\ref{gravmasscomb}))
 the quaternionic prepotential is identified with the $\rSU(2)$ connection, so that, under a  holonomy transformation $U$, it  transforms inhomogeneously as
$$\Gamma \to \Gamma^\prime = U^{-1} \Gamma U +  U^{-1} d U .$$
For a holonomy transformation corresponding to an $\rSU(2)$ SS phase, we have
$$\omega_5^{AB}= \omega_u^{AB}M^u_{\ v} q^v + M^{AB}= \omega_u^{AB}k^u_0 + M^{AB} = P_0^{AB}.$$    
Analogously, for a holonomy tranformation corresponding to a symplectic SS phase, we have
$$\Delta_5^{\alpha\beta}= \Delta_u^{\alpha\beta}M^u_{\ v} q^v + M^{\alpha\beta}= \Delta_u^{\alpha\beta}k^u_0 + M^{\alpha\beta}.$$

The leading terms in the connections ${\Gamma_5}^I_{\ j}$, ${\Delta_5}^\alpha_{\ \beta}$, ${\omega_5}^A_{\ B}$, 
giving the SS phases on ${\mathcal {M}}_\C \times \mathcal{M}_\Q$,
 are 
$M^I_{\ j}$, $M^\alpha_{\ \beta}$, $M^A_{\ B}$ respectively.

On the other hand, the leading terms in the scalar vielbein are
$$ P^{i}_5 \sim M^i_j \varphi^j \,;\quad \mathcal{U}^{\alpha A}_5 \sim M^\alpha_{\ \beta} q^{\beta A} + M^A_{\ B} q^{\alpha B} .$$

The scalar potential is extremized 
at 
$$\mathcal{U}^{\alpha A}_u k^u_0 =0$$
that is for
\begin{equation} \left[ M_{AB} \C_{\alpha\beta} + M_{ \alpha\beta} \epsilon_{AB}
\right] \mathcal{U}^{\beta B}_v q^v =0 .
\label{extremum}
\end{equation}

If $M_{AB}=0$, then the vanishing eigenvalues of $M_{\alpha\beta}$ leave massless some of the  hypermultiplets.
On the other hand, if $M_{AB}\neq 0$, then the extremum is at $q^u=0$. In this case, all the scalars in the hypermultiplets are massive,
thus breaking supersymmetry.


\section{Some illustrative examples}

We are going to consider in this section specific models as  illustrative examples.

For all symmetric real-special manifolds $\mathcal{M}_\R=G_\R/H_\R$, the duality transformations acting linearly on the 
vector potentials belong to 
$$[G_\R \times \rSO(1,1)]{\circledS}~ T^{n_v+1}$$
which is a non-semisimple subgroup of $G_\C$ (the isometry group of the corresponding $D=4$ symmetric space).
Under $G_\R \times \rSO(1,1)$, $G_\C$ decomposes as
$$\fg_\C = \fg_\R \oplus \fso(1,1) + \ft^{n_v+1}_{-2} + \ft '^{n_v+1}_{2}$$
where ${\bf n_v+1}$ is a representation of $G_\R$.
Note that the $D=4$ vectors are in a $2(n_v+2)$-dimensional symplectic representation $\mathcal{R}_\C$ of $G_\C$ which, under $G_\R$, decomposes as follows
$$ \mathcal{R}_\C = \left(\mathcal{R}_\R^{+1} + {\bf 1}^{+3}\right) \oplus  \left(\mathcal{R'}_\R^{-1} + {\bf 1}^{-3}\right). $$
The $(n_v+2)$-dimensional gauge algebra is $\rU(1)\times T^{n_v+1}$, where $\rU(1)$ is a Cartan generator of the maximal compact subgroup $H_\R \subset G_\R$.

 The $\left((n_v+1) \times (n_v+1)\right)$ SS phase matrix has always one vanishing eigenvalue,
because there is always one singlet  of the global symmetry $G_\R$ under its maximal compact subgroup $H_\R$ (corresponding to the $D=5$ graviphoton).
Other vanishing eigenvalues come from the Cartan subalgebra of $H_\R$, depending on the specific symmetric-space under consideration. 

We also give an example of SS phase for a homogeneous non-symmetric special manifold \cite{ale,cec,dvvp}.
Here the number of vanishing eigenvalues of the SS phase 
may be larger than in the symmetric case.

For a generic non-symmetric, non-homogeneous, real-special manifold, the number of vanishing eigenvalues of the SS phase can be rather large, because the 
compact isometry may act on a reduced set of special coordinates.

For the symmetric quaternionic manifolds $G_\Q/[H_\Q\times \rSU(2)_R]$, we find that the SS phase matrix has no vanishing eigenvalues,
for a generic element of the Cartan subalgebra of the maximal compact subgroup $H_\Q \subset G_\Q$ commuting with $\rSU(2)_R$.
If no $\rSU(2)_R$ phase is introduced, then  the SS mechanism give supersymmetric masses to BPS hypermultiplets.
On the other hand, if a $\rSU(2)_R$ phase is switched on, then supersymmetry is broken and a splitting in the masses of the various fields, due to this phase,
occurs in all the multiplets of the supergravity theory. In particular, the splitted fields are the gauginos in the vector multiplet sector, and the quaternionic scalars in the hypermultiplet sector.
The two gravitinos acquire a common mass.

\subsection{Symmetric spaces}
Let us consider the models  based on  symmetric spaces \cite{cvp}.
All the possible $N=2$ $D=4$ symmetric space $\sigma$-models for vector multiplets, which have a $D=5$ counterpart \cite{gst}, are listed in Table 1, while the 
symmetric  quaternionic manifolds \cite{bw} are  indicated in Table 2.

We are going to sketch the various models, giving for each one the mass spectrum of the four dimensional theory, when all the possible SS phases are switched on.

\begin{table}[h]
\begin{center}
\begin{tabular} {|c|c|}
\cline{1-2}  $\mathcal{M}_{\R}$ & $\mathcal{M}_{\C}$
\\ \cline{1-2}
\hline
{\em SG} & $\frac{\rSU(1,1)}{\rU(1)}$ 
\\ 
$\frac{\rSL(3,\R)}{\rSO(3)}$&$\frac{\rSp(6,\R)}{\rU(3)}$\\
$\frac{\rSL(3,\C)}{\rSU(3})$&$\frac{\rU(3,3)}{\rU(3)\times\rU(3)}$ \\
 $\frac{\rSU^*(6)}{\rUSp(6)}$&$\frac{\rSO^*(12)}{\rU(6)}$ \\
 $\frac{\rE_{6,-26}}{\rF_4}$&$\frac{\rE_{7,-25}}{\rE_6\times\rSO(2)}$ \\
 $\rSO(1,1) \times   \frac{\rSO(1,n_v-1)}{ \rSO(n_v-1)}$ & $\frac{\rSU(1,1)}{\rU(1)}\times \frac{\rSO(2,n_v)}{\rSO(2) \times \rSO(n_v)}$ 
\\
\cline{1-2}
\end{tabular}
\caption{All the symmetric cosets participating to $N=2$ SS mechanism in the vector multiplet sector:
the real-special 
manifolds in $D=5$ ($\mathcal{M}_{\R}$) and the corresponding very-special 
manifolds in $D=4$ ($\mathcal{M}_{\C}$).}
\label{examples}
\end{center}
\end{table}

\subsubsection{Vector multiplet sector}
\begin{itemize}
\item{
The first case is quite degenarate, since it corresponds to
$\mathcal{M}_{\R}= 0$ (pure supergravity); $\mathcal{M}_{\C}=\frac{\rSU(1,1)}{\rU(1)}$;
 dim$_\C \left(\mathcal{M}_{\C}\right)=1$.\\
In this case one may still build a SS phase in the Cartan subalgebra (CSA) of the global R-symmetry  $\rSU(2)_R$.\\
As discussed in Section 3, this correspond to having a Fayet--Iliopoulos term in $D=4$, which breaks supersymmetry.
In this case, which is the $N=2$ analogue of the no-scale model of \cite{cfkn}, the scalar potential is identically zero.

All fermions take a mass equal to the phase, while the bosons stay massless.
A situation like the one described for this model, with SS phase in the R-symmetry breaking supersymmetry,  
may also occur for all the other models listed in this subsection, and will not be repeated.  
}
\item{
$\mathcal{M}_{\R}=\frac{\rSL(3,\R)}{\rSO(3)} \subset \frac{\rSp(6,\R)}{\rU(3)}=\mathcal{M}_{\C}$;
dim$_\R \left(\mathcal{M}_{\R}\right)=5$, dim$_\C \left(\mathcal{M}_{\C}\right)=6$.\\
SS phase: $m \in$ CSA of $\rSO(3)$.\\
In $D=5$: 6 vectors in the two-fold symmetric representation of $\rSL(3,\R)$.
${\bf 6} \to {\bf 5}+{\bf 1} $ under $\rSO(3)$.
\\
Mass eigenvalues: $(-2m,-m,0,m,2m)$.

$D=4$ spectrum of 6 vector multiplets + one graviphoton:

2 massless vector multiplets\\
1 massive BPS vector multiplet, mass $|m|$\\
1 massive BPS vector multiplet, mass $|2m|$\\
1 massless graviphoton.
}
\item{
$\mathcal{M}_{\R}=\frac{\rSL(3,\C)}{\rSU(3)} \subset \frac{\rU(3,3)}{\rU(3)\times \rU(3)}=\mathcal{M}_{\C}$;
dim$_\R \left(\mathcal{M}_{\R}\right)=8$, dim$_\C \left(\mathcal{M}_{\C}\right)=9$.\\
SS phases: $(m_1,m_2) \in$ CSA of $\rSU(3)$.\\
In $D=5$: 9 vectors in the  $({\bf 3},{\bf \bar 3}) \in \rSL(3,\C)$.\\
$({\bf 3},{\bf \bar 3}) \to {\bf 8}+{\bf 1} $ under $\rSU(3)$.
\\
$D=4$ spectrum of 9 vector multiplets + one graviphoton:

3 massless vector multiplets\\
2 massive BPS vector multiplets, mass $|m_1 \pm 3 m_2|$\\
1 massive BPS vector multiplet, mass $|2 m_1 |$\\
1 massless graviphoton.
}
\item{
$\mathcal{M}_{\R}=\frac{\rSU^*(6)}{\rUSp(6)} \subset \frac{\rSO^*((12)}{\rU(6)}=\mathcal{M}_{\C}$;
dim$_\R \left(\mathcal{M}_{\R}\right)=14$, dim$_\C \left(\mathcal{M}_{\C}\right)=15$.\\
SS phases: $(m_1,m_2,m_3) \in$ CSA  of $\rUSp(6)$\\
In $D=5$: 15 vectors in the  two-fold antisymmetric rep. of $\rSU^*(6)$.\\
${\bf 15} \to {\bf 14}+{\bf 1} $ under $\rUSp(6)$.
\\
$D=4$ spectrum of 15 vector multiplets + one graviphoton:

3 massless vector multiplets\\
6 massive BPS vector multiplets, mass $|m_i \pm m_j|$, ($i<j=1,2,3$)\\
1 massless graviphoton.
}
\item{
$\mathcal{M}_{\R}=\frac{\rE_{6(-26)}}{\rF_4} \subset \frac{\rE_{7(-25)}}{\rU(1)\times E_6}=\mathcal{M}_{\C}$;
dim$_\R \left(\mathcal{M}_{\R}\right)=26$, dim$_\C \left(\mathcal{M}_{\C}\right)=27$.\\
SS phases: $(m_1,m_2,m_3,m_4) \in$ CSA  of $\rF_4$.\\
In $D=5$: 27 vectors in the  two-fold antisymmetric rep. of $\rE_{6(-26)}$.
${\bf 27} \to {\bf 26}+{\bf 1} $ under $\rF_4$.
\\
$D=4$ spectrum of 27 vector multiplets + one graviphoton:

3 massless vector multiplets\\
8 massive BPS vector multiplets, mass $|m_1 \pm m_2 \pm m_3 \pm m_4 |$,\\
4 massive BPS vector multiplets, mass $|2 m_i |$, ($i=1,\cdots , 4$)\\
1 massless graviphoton.
}
\item{
$\mathcal{M}_{\R}=\rSO(1,1) \times   \frac{\rSO(1,n_v-1)}{ \rSO(n_v-1)}
\subset \frac{\rSU(1,1)}{\rU(1)}\times \frac{\rSO(2,n_v)}{\rSO(2) \times \rSO(n_v)}=\mathcal{M}_{\C}$;\\
dim$_\R \left(\mathcal{M}_{\R}\right)=n_v$, dim$_\C \left(\mathcal{M}_{\C}\right)=n_v+1$.\\
SS phases: $(m_1,\cdots ,m_{[\frac{n_v-1}{2}]}) \in$ CSA  of $\rSO(n_v -1)$.\\
In $D=5$ we have $n_v+1$ vectors. However,
in this case the $D=5$ graviphoton together with $n_v-1$ vectors of the matter multiplets compose the vector representation of $\rSO(1,n_v-1)$, 
while 1 vector multiplet is inert, so that\\
${\bf n_v}+{\bf 1}  \to  {\bf (n_v -1)}+{\bf 1}+{\bf 1} $ under $\rSO(n_v-1)$.\\
There are always 2 singlets, corresponding to 2 massless vectors; then 2 different cases:
\begin{itemize}
\item{ $n_v$ even\\
In this case the $(n_v-1)$-vector has one vanishing eigenvalue so that we then have:\\
3 massless vector multiplets\\
$(n_v-2)/2$ massive BPS vector multiplets, mass $| m_i |$, ($i=1,\cdots ,\frac{n_v-2}{2} $)\\
1 massless graviphoton.
}
\item{$n_v$ odd:\\
In this case the $(n_v-1)$-vector has no vanishing eigenvalues so that we then have:\\
2 massless vector multiplets\\
$(n_v-1)/2$ massive BPS vector multiplets, mass $| m_i |$, ($i,..=1,\cdots ,\frac{n_v-1}2 $)\\
1 massless graviphoton.}
\end{itemize}
}
\end{itemize}
To summarize, looking to the $D=5$ SS compactifications which give $D=4$ symmetric spaces
 we find in two cases a spectrum with only two massless multiplets left (the rest being 
massive BPS Higgs supermultiplets), while in all the other cases three vector multiplets stay massless.

\subsubsection{The hypermultiplet sector}
We have listed in Table 2 all the symmetric-space quaternionic manifolds. Each of them could in principle participate to the SS mechanism, since they fulfill the
 requirement of having a compact isometry which may be gauged by the graviphoton.\footnote{
Note that the same observation is true for all the quaternionic manifolds (not necessarily
cosets) which are obtainable as the c-map \cite{cfg} of some special Kahler manifold, since the c-map construction implies that a $\rU(1) \subset \rSL(2,\R)$ isometry
is always present on the final quaternionic manifold. We are not going to discuss these more general cases here.}

\begin{table}[h]
\begin{center}
\begin{tabular} {|c|c|}
\hline   $\mathcal{M}_{\Q}$ & dim$_\Q(\mathcal{M}_{\Q})$
\\ \hline
 $\frac{\rG_{2(+2)}}{SO(4)}$ & 2
\\ 
 $\frac{\rF_{4(+4)} }{\rUSp(6) \times \rSU(2)}$ & 7\\
 $\frac{\rE_{6(+2)}}{ \rSU(6) \times \rSU(2)}$ & 10 \\
$\frac{E_{7(-5)} }{\rSO(12) \times \rSU(2)}$ & 16 \\
$\frac{\rE_{8(-24)} }{\rE_7 \times \rSU(2)}$ & 28
\\ 
$\frac{\rSO(4,n)}{\rSO(4) \times \rSO(n)}$ & $n$\\
 $\frac{\rU(2,n)}{\rU(2) \times \rU(n)}$ &$n$\\
 $\frac{\rUSp(2,2n)}{\rSU(2) \times \rUSp(2n)}$&$n$\\
\hline
\end{tabular}
\caption{All the quaternionic symmetric coset-manifolds participating to $N=2$ SS mechanism. In particular, the first 
6 are the  c-map of the very-special manifolds of Table 1.}\label{examples2}
\end{center}
\end{table}

For all the models we find that, in the general case where all the SS phases in the symplectic part of the isometry are switched on, all the hypermultiplets become massive
BPS multiplets.

All the quaternionic scalars are also charged with respect to the R-symmetry $\rSU(2)_R$ (in the fundamental representation).
So, by introducing the corresponding SS phase $m$, supersymmetry is  broken, giving an extra mass $\pm m$ to all the scalars in the hypermultiplets.
We are not going to discuss further the $\rSU(2)_R$ phase $m$, confining the discussion in the sequel to the SS phase in the $\rUSp(2n_h)$ part.
In detail we find
\begin{itemize}
\item{
$\mathcal{M}_\Q=\frac{\rG_{2(2)}}{\rSO(4)}$;
dim $\left(\mathcal{M}_\Q\right)=8$; $n_h =2$. \\
SS phase: $q$ in  the CSA of $\rSU(2)$.\\
$8 \to \left(4 ,2 \right) $ under $\rSU(2)\times \rSU(2)_R$.
\\
 No state is neutral under $\rSU(2)$, so
 the 2 multiplets become BPS, one  with mass  $q$ and the other with mass $2q$.
}
\item{
$\mathcal{M}_{\Q}=\frac{\rF_{4(4)}}{\rUSp(6)\times \rSU(2)_R}$; dim$\left(\mathcal{M}_{\Q}\right)=28$; $n_h =7$.  \\
SS phase: $(q_1,q_2,q_3)$ in the CSA of $\rUSp(6)$.\\
${\bf 28} \to \left(14' , 2\right) $ under $\rUSp(6)\times \rSU(2)$ ($14'$: three-fold antisymmetric representation of $\rUSp(6)$).
\\
No state is neutral under $\rUSp(6)$; the BPS spectrum is:\\
4 BPS multiplets with mass $|q_1\pm q_2\pm q_3|$\\
3 BPS multiplets with mass $|q_i|$ ($i=1,2,3$)\\
}
\item{
$\mathcal{M}_{\Q}=\frac{\rE_{6(2)}}{\rSU(6)\times \rSU(2)_R}$; dim$\left(\mathcal{M}_{\Q}\right)=40$; $n_h =10$.  \\
SS phase: $(q_1,\cdots q_5)$ in the CSA of $\rSU(6)$.\\
$ 40 \to \left(20 , 2\right) $ under $\rSU(6)\times \rSU(2)_R$ (20: three-fold antisymmetric representation of $\rSU(6)$).
\\
No state is neutral under $\rSU(6)$; the BPS spectrum is:\\
10 BPS multiplets.
}
\item{
$\mathcal{M}_{\Q}=\frac{\rE_{7(-5)}}{\rSO(12)\times \rSU(2)_R}$; dim$\left(\mathcal{M}_{\Q}\right)=64$; $n_h =16$.  \\
SS phase: CSA of $\rSO(12)$.\\
$ 64 \to \left(32 , 2\right) $ under $\rSO(12)\times \rSU(2)_R$ (32: spinor representation of $\rSO(12)$).
\\
No state is neutral under $\rSO(12)$.\\
16 BPS multiplets.
}
\item{
$\mathcal{M}_{\Q}=\frac{\rE_{8(-24)} }{\rE_7 \times \rSU(2)_R}$; dim$\left(\mathcal{M}_{\Q}\right)=112$; $n_h =28$.  \\
SS phase: CSA of $\rE_7 $.\\
$ 112  \to \left(56 , 2\right) $ under $\rE_7\times \rSU(2)_R$ (56: fundamental representation of $\rE_7 $).
\\
No state is neutral under $\rE_7$.  \\
28 BPS multiplets.
}
\item{
$\mathcal{M}_{\Q}=\frac{\rSO(4,n)}{\rSO(4) \times \rSO(n)}$; dim$\left(\mathcal{M}_{\Q}\right)=4n$; $n_h =n$.  \\
SS phase: CSA of $\rSU(2) \times \rSO(n)$.\\
$ 4n  \to \left(2,n , 2\right) $ under $\rSU(2) \times \rSO(n) \times \rSU(2)_R$. 
\\
No state is neutral under $\rSU(2) \times \rSO(n) \subset \rUSp(2n)$\\
the generic configuration has all massive hypermultiplets.\\
If we set a phase only with respect to $\rSU(2)_D$, then, since ${\bf 2} \times {\bf 2} = {\bf 3} + {\bf 1}$ we have two vanishing eigenvalues 
in this $\rSU(2)_D$ phase; if $n$ is odd, we have one zero-eigenvalue also in the $\rSO(n)$ phase. In this case we finally have 2 massless hypermultiplet scalars, 
but supersymmetry is broken. 
}
\item{
$\mathcal{M}_{\Q}=\frac{\rU(2,n)}{\rU(2) \times \rU(n)}$; dim$\left(\mathcal{M}_{\Q}\right)=4n$; $n_h =n$.  \\
SS phase: CSA of $\rU(1) \times \rU(n)$.\\
$ 4n  \to \left(n , 2\right)_+ + \left(\bar n , 2\right)_- $ under $(\rU(n) \times \rSU(2)_R)_{\rU(1)}$. 
\\
No state is neutral under $\rU(1) \times \rU(n) \subset \rUSp(2n)$.\\
Analogously to the case above, in the generic configuration all  hypermultiplets are massive.
}
\item{
$\mathcal{M}_{\Q}=\frac{\rUSp(2,2n)}{\rSU(2)_R \times \rUSp(2n)}$; dim$\left(\mathcal{M}_{\Q}\right)=4n$; $n_h =n$.  \\
SS phase: CSA of $\rUSp(2n)$.\\
$ 4n  \to \left(2n , 2\right) $ under $(\rUSp(2n) \times \rSU(2)_R)$.
\\
No state is neutral under $\rUSp(2n)$.\\
As for the cases above, in the generic configuration all  hypermultiplets are massive.
}
\end{itemize}

\subsection{Non symmetric spaces}

For non-symmetric spaces we can still consider SS phases if there are some compact isometries on the manifold.
This is generally possible for the case of homogeneous real geometries of $D=5$, $N=2$ scalar manifolds, that have been classified in the literature \cite{dwvp2}
and denoted by $L(q,P,\dot P)$. The compact isometries of these real manifolds are $\rSO(q+1) \times \mathcal{S}_q(P,\dot P)$, where 
$ \mathcal{S}_q(P,\dot P)$ is the metric preserving group in the centralizer of the Clifford algebra $\mathcal{C}(q+1,0)$ \cite{dvvp}.

Let us for instance consider the space $L(0,P,\dot P)$, with compact isometry group $\rSO(P)\times \rSO(\dot P)$.
Here $n_v=2+P+\dot P$.
In the  $D=4$ theory, if $P,\dot P$ are both even, we have three massless vector multiplets, if one is even and the other odd, we have 4 massless vector multiplets, while if 
$P,\dot P$ are both odd we have 5 massless vector multiplets, all the rest being $\frac 12$-BPS multiplets..

A similar discussion can be made for homogeneous, non-symmetric, quaternionic spaces which are obtained as c-map of d-geometries. They
have the compact group of isometries
 $\rSO(q+3)\times \rSO(3)\times \mathcal{S}_q(P,\dot P)$ \cite{dvvp}, and one may again introduce
suitable SS phases in this sector.


\section*{Acknowledgements}
M. A. Ll. wants to thank the Theory Division at CERN for its
hospitality during the realization of this work.

 The work of S.F. has been supported in
part by the D.O.E. grant DE-FG03-91ER40662, Task C, and in part by
the European Community's Human Potential Program under contract
HPRN-CT-2000-00131 Quantum Space-Time, in association with INFN
Frascati National Laboratories.

The work of M. A. Ll. has been supported by the
research grant BFM 2002-03681 from the Ministerio de Ciencia y
Tecnolog\'{\i}a (Spain) and from EU FEDER funds.

\appendix


\section{Glossary of very-special geometry and\\ SS Killing prepotentials}

In the $z^I$ four-dimensional variables the special coordinates are
$$X^\Lambda =(z^0 =1 , z^I ) \qquad L^\Lambda =(e^{\frac K 2} , e^{\frac K 2}z^I )
$$
therefore
$$f^\Lambda_i = e^{\frac K 2} (\partial_i + K_{,i}) X^\Lambda =\left( e^{\frac K 2} (\delta^I_i +   K_{,i}X^I),e^{\frac K 2}  K_{,i}\right)$$
where $I, i$ here  are flat and world indices respectively on the very-special manifold.
Moreover, for the Killing vectors one has that
$$ k^I_\Lambda L^\Lambda = k^I_0 L^0 + k^I_J L^J =0$$
is satisfied for 
$$k^I_J = -M^I_{\ J} \qquad k^I_0 = M^I_{\ J} z^J$$
which implies the identity, on the corresponding prepotential
$$P_\Lambda L^\Lambda = P_0 L^0 + P_I L^I = e^{\frac K 2} \left( K_{,I} M^I_{\ J}z^J - K_{,I} M^I_{\ K}z^K \right) =0$$
and analogously
$$ P_\Lambda \bar L^\Lambda =0$$

In the $z^I$ variables, the K\"ahler potential is invariant under $\rU(1)$ and  Peccei--Quinn symmetries 
$$\delta z^I =   M^I_{\ J}z^J \xi^0- M^I_{\ J}\xi^J$$
so that 
$$\delta_{\xi_I} K =  K_{,I}   M^I_{\ J}\xi^J + K_{,{\bar I}}   M^I_{\ J}\xi^J =0$$
since
$ K_{,I}=- K_{,{\bar I}}$.
Correspondingly, for the $\xi^0$ transformation
$$\delta_{\xi_0} K = \left( K_{,I}   M^I_{\ J}z^J + K_{,{\bar I}}   M^I_{\ J} \bar z^J\right) \xi^0 =0$$
which gives
$$ K_{,I}   M^I_{\ J}\Im z^J =0$$
This is because $M^I_{\ J}$ is an isometry of the very-special K\"ahler manifold which has a linear action on the $z^J$ coordinates.

The prepotential $P_\Lambda$ is therefore given by 
$$P_0 = K_{,I}   M^I_{\ J}z^J \, , \quad P_J = - K_{,I}   M^I_{\ J}$$
so that 
$$k^K_0 = K^{,K\bar I} \partial_{\bar I} P^0 = K^{,K\bar I} K_{,I\bar I}   M^I_{\ J}z^J = M^I_{\ J}z^J . $$

We note that, in the case of symmetric spaces $G_\C /[\rU(1) \times H_\C]$ based on d-geometries, the compact symmetry $H_\C$ is not manifest (non linearly realized) 
on  the chosen special coordinates $z^I$.
The non-compact symmetry $G_\R$  is manifest, and so is the compact symmetry $H_\R \subset H_\C$.

This fact is already understood for the $[\frac{\rSU(1,1)}{\rU(1)} \times \frac{\rSO(2,n)}{\rSO(2)\times\rSO(n)}]$ series.
In this case, for the cubic prepotential it is necessary to choose a symplectic basis where only $\rSO(1,n-1)\times \rSO(1,1) \subset \rSO(2,n)$
is manifest \cite{ckpdfwg}. For a choice of special coordinates which makes all $\rSO(2,n)$ manifest, a prepotential function does not exist \cite{cdfv}.


\section{Relation between real-special geometry in $D=5$ and very-special geometry in $D=4$}

The relation between the geometries of the scalar manifolds of vector multiplets in five and four dimensions comes by inspection 
of the bosonic kinetic terms in the lagrangian, in the $D=4$ Einstein frame, as they come by dimensional reduction from five dimensions \cite{gst,ccdf,cfgk}
\begin{eqnarray}
\frac 1{\sqrt{-g_{(4)}}} \mathcal{L}_{(4)}&=& -\frac 12  \mathcal{R}_{(4)}-\frac 12 G_{IJ}(t) \partial_\mu t^I \partial^\mu t^J
-\frac 1{12} \partial_\mu  \log \left(\phi_5\right)^3  \partial^\mu  \log\left( \phi_5\right)^3 \nonumber\\
&&- \frac 12 G_{IJ}(t)\left( \phi_5\right)^{-2} \partial_{\mu}A^I_5  \partial^\mu A^J_5  - \frac 14 G_{IJ}(t)~  \phi_5 F^I_{\mu\nu} F^{J ~ \mu\nu} \label{lag4}
\end{eqnarray}
where $\phi_5=e^\sigma$ is the 5th component of the vielbein ($S^1$ radius).
We have not reported in (\ref{lag4}) the contributions from the KK graviphoton and from the 5D Chern--Simons term (see \cite{gst}), which are not relevant for the present discussion.

Here
$$G_{IJ} = -\frac 12 \partial_I \partial_J \log \mathcal{V}|_{\mathcal{V}=1}$$
where $\mathcal{V}\equiv d_{IJK} t^I t^J t^K$, with $t^I=t^I(\varphi^i)$, $i=1,\cdots n_v$ and $I=1,\cdots n_v+1$.

The surface $\mathcal{V}=1$ defines the $n_v$-dimensional real manifold $\mathcal{M}_\R^{n_v}$ spanned by the scalars in the vector multiplets 
at $D=5$, describing real-special geometry. 

Explicitly we have
$$G_{IJ} = -3 t_{IJ} + \frac 92 t_I t_J$$
where $t_I \equiv d_{IJK}t^J t^K$, $t_{IJ} \equiv d_{IJK} t^K$.

the metric on $\mathcal{M}^\R_{n_v}$ is
$$g_{ij} = G_{IJ}t^I_{,i} t^J_{,j}\quad\mbox{ with }\quad t^I_{,i}=\frac{\partial t^I}{\partial \varphi^i}.$$

Note that $t^It_{I,i}=t_It^I_{,i}=0$, and also
$$g_{ij} = -3d_{IJK}t^I_{,i} t^J_{,j}t^K
= -\frac 32 t_{I,i}t^I_{,j}.$$

If we now introduce $n_v+1$ variables $v^I$ defined by
$v^I = \phi_5 t^I$, so that 
$d_{IJK}v^I v^J v^K = \left(\phi_5\right)^3=\mathcal{V}$,
from the lagrangian (\ref{lag4}) it immeditely follows
\begin{eqnarray}
\frac 1{\sqrt{-g_{(4)}}} \mathcal{L}_{(4)}&=& -\frac 12  \mathcal{R}_{(4)}-\frac 12 G_{IJ}(v) \partial_\mu v^I \partial^\mu v^J
\nonumber\\
&&- \frac 12 G_{IJ}(t)\left( \phi_5\right)^{-2} \partial_{\mu}A^I_5  \partial^\mu A^J_5  - \frac 14 G_{IJ}(t)~  \phi_5 F^I_{\mu\nu} F^{J ~ \mu\nu} \label{lag4'}
\end{eqnarray}
where 
\begin{equation}
G_{IJ}(v)= -\frac 12 \frac{\partial}{\partial v^I}  \frac{\partial}{\partial v^J} \log \mathcal{V}= \left(\phi_5\right)^{-2} G_{IJ}(t).
\label{metr}
\end{equation}
By combining the axions $a^I = A^I_5$ with the real scalars $v^I$ to form complex variables
$$z^I = -a^I + {\rm i} v^I,$$
the scalar kinetic term of the above lagrangian can be rewritten in the more compact form
$$- g_{I\bar J}\partial_\mu z^I \partial^\mu \bar z^{\bar J}$$
with $
g_{I\bar J} \equiv -\partial_I \partial_{\bar J} \log \mathcal{V} = \frac 12 G_{IJ}
$ the very-special-K\"ahler metric.

We can now set the relation between $5D$ and $4D$ special coordinates.
We have, in $D=4$
$$L^\Lambda = e^{\frac K2} X^\Lambda \, ,\qquad X^\Lambda =\left(X^0,X^I\right)$$
with $X^0=1$, $X^I=z^I$.

This choice identifies the four-dimensional graviphoton with the \KK vector $B_\mu$, in accordance with the gauged supergravity
interpretation of the SS reduction.

The $n_v+2$ vector fields are $\left(A^0_\mu = B_\mu\,; Z^I_\mu = A^I_\mu - A^I_5 B_\mu\right)$ and the structure constants of the flat group are
$f_{\Lambda \Sigma}^{\ \ \Delta} = f_{I0}^{\ \ J}$ (zero otherwise).
 The $B_\mu$ gauge field gauges a $\rU(1)$ isometry which may have components 
both on the very-special manifold and on the quaternionic manifold
$$B=I^{SG} + I^Q_{\rUSp(2n_h)} + I^Q_{\rSU(2)}.$$ 

\bigskip

The contribution to the scalar potential in $D=4$ from the real-special geometry isometries also comes from dimensional reduction of the kinetic term of the scalars 
in the vector multiplets, through the term, in the dimensionally reduced lagrangian
$$\frac 12 e^{-3\sigma} P^a_5 P_{a 5}.$$
By using the fact that $P^a_5$, the 5th component of the pull-back on space-time of the real-special geometry vielbein, is
$P^a_5 = P^a_{,i}\partial_5 \varphi^i = P^a_{,i} k^a_0$, we then obtain 
$$ P^a_5 P_{a 5}= g_{ij} k^i_0 k^j_0$$
where we used the fact  \cite{cfgk} that
$g_{ij}=P^a_{,i}P_{a,j}=-\frac32 t^I_{,i} t_{I,j}$ and $t_{I,i} =\partial_i t_I = 2 d_{IJK} t^J t^K_{,i}$.
Then, recalling (\ref{kahlerphi}), the contribution to the scalar potential is
$$V_{SG}=\frac 12  e^K g_{ij}k^i_0k^j_0$$
or, in terms of the complex variables $z^I= - a^I_5 +{\rm i} \phi_5 t^I$ and of the very-special K\"ahler metric (\ref{metr})
$$V_{SG}= \frac 12 e^K G_{IJ}k^I_0k^{\bar  J}_0 = e^K g_{I\bar J}k^I_0k^{\bar J}_0,$$
where $k^I_0 \equiv -{\rm i}t^I_{,i} k^i_0= - k^{\bar I}_0$,
which is the $N=2$ contribution from the gauged vector multiplets in $D=4$ (see eqs. (\ref{smkilling}) and (\ref{sgpot})).
We note that this formula agrees with (\ref{sgpot}) if we send $M\to 2M$ in the SS phases.

\end{document}